\documentclass[aps,prd]{revtex4}

\usepackage{amsthm,amsfonts,amsmath,latexsym,amssymb}
\usepackage[american]{babel}
\usepackage[dvips]{graphicx}
\usepackage{mathrsfs}

\newcommand\bea{\begin{eqnarray}}
\newcommand\eea{\end{eqnarray}}

\begin{document}

\title{Zeta Determinant for Laplace Operators on Riemann Caps}

\author{Antonino Flachi}
\email{flachi@yukawa.kyoto-u.ac.jp}
\affiliation{Yukawa Institute for Theoretical Physics, Kyoto University, Kyoto, Japan}

\author{Guglielmo Fucci}
\email{Guglielmo_Fucci@Baylor.edu}
\affiliation{Department of Mathematics, Baylor University, Waco, TX 76798 USA}

\preprint{YITP-10-23}

\date{\today}

\begin{abstract}
The goal of this paper is to compute the zeta function determinant for the massive Laplacian on {\it Riemann caps} (or {\it spherical suspensions}). These manifolds are defined as compact and boundaryless $D-$dimensional manifolds deformed by a singular Riemannian structure. The deformed spheres, considered previously in the literature, belong to this class. After presenting the geometry and discussing the spectrum of the Laplacian, we illustrate a method to compute its zeta regularized determinant. The special case of the deformed sphere is recovered as a limit of our general formulas.
\end{abstract}
\maketitle

\section{Introduction}
\label{sec1}

Zeta functions are a topic of ubiquitous interest due to their wide range of applicability. The basic reason for their popularity is that systems with boundaries or finite size effects (in general boundary conditions) can be conveniently studied by constructing and analyzing suitable functions of the spectrum of some operator.
For instance, many fundamental properties of quantum fields are encoded in the effective action, that contains information about the induced quantum energy-momentum tensor that is the fundamental object necessary to discuss, for instance, how the classical equations of motion are modified by quantum effects, or phenomena like particle creation in curved space. The effective action $\Gamma$ can be expressed as
\bea
\Gamma \sim \ln \det {\cal O}~,
\eea
with $\cal O$ being (typically) a positive elliptic differential operator. When, the operator $\cal O$ acts over
a compact manifold, the spectrum is also discrete. The effective action is then formally written as sum over the eigenvalues (the sum is, in fact, an integral with measure depending on the spectral density of the eigenvalues in the continuous case)
\bea
\Gamma \sim \sum_{\lambda} \ln \lambda~.
\eea
Written as above, the effective action is divergent and it does not have a precise meaning unless it is regularized by using some appropriate analytic continuation procedure. One such procedure, introduced in \cite{RS,DC,HWK}, uses the zeta function
\bea
\zeta(s) = \sum_{\lambda} \lambda^{-s}~,
\eea
to define the effective action as the analytic continuation of the derivative $\zeta'(s)$ at $s=0$. This is, in essense, the idea behind zeta function regularization and the standard textbook expression for the effective action is \cite{parkertoms}:
\bea
\Gamma = -{1\over 2} \zeta'(0) -{1\over 2} \zeta(0)\ln \mu^2~,
\eea
where $\mu$ is a renormalization scale. Sometimes the above functional determinant is also addressed as zeta regularized determinant.

Finding the zeta function and performing the analytic continuation is, most of the times, a non-trivial task, and only a few general methods are available. When the spectrum can be found explicitly (this is typically possible when the background manifold is highly symmetric and the differential operator in question is conformally covariant), aside from trivial cases, different ways have been adopted to compute the zeta function and functional determinant. A classical example is that of the Laplacian on a spherical domain analyzed in
ref.~\cite{dowker}. In more general set-ups, when eigenvalues are known only implicitly, one may use the method developed in \cite{bordag96}.

Up to now, explicit results for the zeta determinant of the Laplacian have been obtained in a variety of cases with different methods that, aside of those mentioned of spheres \cite{dowker} and orbifolded spheres \cite{chang}, include projective spaces \cite{spreaficoosp}, balls \cite{bordag96}, hyperbolic manifolds \cite{camporesi}, cones \cite{dowker,bordag}, intervals \cite{forman}.

One interesting case, beyond those that have been analyzed so far, is that of the deformed spheres, the simplest example of such geometries being a spherical cap. The deformed spheres can also be seen as generalization of lunes and orbifolded spheres. These type of geometries occur frequently in cosmology, since sectors of de Sitter space belong to this class. In general, a deformed sphere is defined as a $D-$dimensional sphere equipped with a singular Riemannian structure. The metric is written as
\bea
ds_D^2 = d\theta^2 + \sin^2 \theta ds_{D-1}^2~,
\eea
where $ds_{D-1}^2$ is the metric of a $(D-1)-$dimensional sphere. Ref.~\cite{spreafico} describes the geometry of deformed sphere and we address the reader to that reference for details.
The zeta determinant of the Laplacian on the above manifolds has been recently analyzed in Ref.~\cite{spreafico} that introduced a general method to deal with some classes of simple and double abstract zeta functions allowing to compute the zeta invariants $\zeta(0)$ and $\zeta'(0)$, and presented explicit formulas in $2$ and $3$ dimensions. Alternative computations include those of ref.~\cite{dowker} that considered the case of spherical domains, spherical caps and lunes and computed the functional determinant in a variety of dimensions. Some related work in physics, aiming at computing the effective action in the context of quantum cosmology \cite{barvinsky}, discussed the analogous problem of computing the zeta determinant for a sector of four dimensional de Sitter space. In brane world cosmology, ref.~\cite{flachi1} dealt with the computation of the zeta determinant in $5-$ dimensional anti-de Sitter space bounded by de sitter branes and in sectors of higher dimensional de Sitter \cite{flachi2}. Finally, the Casimir energy for two-dimensional deformed spheres has been considered in ref.~\cite{svc}. 

In this work, we will focus our attention on some generalizations of the deformed spheres in which the base manifold ${\mathscr{N}}$ is smooth, compact and without boundary, but otherwise unspecified. The base manifold is deformed by a singular Riemannian structure analogous to that of the deformed spheres and the local geometrical properties of the spherical suspension can be described by use of the following metric
\begin{equation}
ds^2=d\theta^2 +\sin^2\theta d\Sigma^2~,
        \label{eq1:metric}
\end{equation}
where $d\Sigma^2$ represents the line element of the base manifold ${\mathscr{N}}$. The angular coordinate $\theta$ is restricted to the interval $\left[0,\theta_0\right]$ with $\theta_0\in(0,\pi)$. We will indicate this manifold as $\Omega$ and use the terminology of {\it Riemann Cap}. Another commonly used name is {\it spherical suspension}.

The goal of the paper is to present a method to compute the functional determinant for the Laplacian on $\Omega$. By confronting the problem with that studied in ref.~\cite{bordag}, one immediately notices that one needs to consider the asymptotic behavior of new eigensolutions. This is not a merely technical complication and, as explicit computations show, the method developed in \cite{bordag} does not easily generalize to this case, although, in principle, directly applicable.
Here, we take a different route, analogous to that developed in ref.~\cite{barvinsky}, that makes use of the direct analytic continuation at $s=0$ of the integral representation of the zeta function. The method is valid for only when the (uniform) asymptotic expansion of the eigenfunction satisfies a certain form, but it is general enough to encompass the cases treated by the method of \cite{bordag} and also those of the spherical suspension.
One of the aims of the paper is precisely to refine the results of ref.~\cite{barvinsky} in order to be able to treat the case of Riemann caps.

This paper is divided in four main sections. Sec. \ref{sec2} is devoted to describe spectrum, eigenvalues and eigenfunctions, of the Laplacian on $\Omega$. In sec. \ref{sec3} we present the general method used to compute the functional determinant. The general results for the regularized zeta determinant of the Laplacian on $\Omega$ are reported in sec. \ref{sec4}. The case of the deformed spheres, considered in \cite{spreafico}, is discussed in sec. \ref{sec5} and previous results recovered from our general formulas. The final section is devoted to summarize and discuss the results of our work.

\section{Geometry and Spectrum}
\label{sec2}
The Laplacian on $\Omega$ can be explicitly written, in spherical coordinates, as
\begin{equation}
\nabla_\Omega =
{\partial^2 \over \partial \theta^2}
+d \cot \theta {\partial\over \partial \theta}
+{1\over \sin^2\theta}\nabla_\Sigma~,
        \label{eq2:laplacian}
\end{equation}
with $\nabla_\Sigma$ being the Laplacian on the base manifold ${\mathscr{N}}$.
The eigenvalue equation is
\begin{equation}
\left[{\partial^2 \over \partial \theta^2}
+d \cot \theta {\partial \over \partial \theta}
+{1\over \sin^2\theta}\nabla_\Sigma +m^{2}\right]\varphi_{n,{\bf k}}(\theta,X_i) = \alpha_{n,{\bf k}}^2 \varphi_{n,{\bf k}}(\theta,X_i)~,
        \label{eq3:eigen_eq}
\end{equation}
where the coordinates on ${\mathscr{N}}$ are labelled by $X_i$, with $i=1,\cdots, d$. The indices $n$ and $\bf k$ have been introduced to distinguish between the angular direction and the base manifold and are, for the moment, unconstrained. The quantity $m$ is constant and represents a mass term once equation (\ref{eq3:eigen_eq}) is interpreted as the Klein-Gordon equation for a massive scalar field propagating on $\Omega$. We assume here that a complete and orthonormal set of of solutions to the eigenvalue equation on ${\mathscr{N}}$ exists, but leave the specific properties of the base manifold and the boundary conditions unspecified and carry out our analysis in general. When the base manifold is a sphere, we reproduce, using a different method, the results of refs.~\cite{spreafico,dowker, flachi2}.
A general solution to (\ref{eq3:eigen_eq}) can be written as
\begin{equation}
\varphi_{n,{\bf k}}(\theta,X_i) =
\phi_{n,{\bf k}}(\theta) {\cal H}_{{\bf k}}(X_i)~,
\end{equation}
where ${\cal H}_{{\bf k}}(X_i)$ are the harmonics on ${\mathscr{N}}$ with degeneracy $d({\bf k})$ and eigenvalues $\lambda_{\bf k}$ satisfying the following equation
\begin{equation}
\nabla_\Sigma {\cal H}_{{\bf k}}(X_i)=
-\lambda_{\bf k}^2
{\cal H}_{{\bf k}}(X_i)~,
        \label{eq5:eqbase}
\end{equation}
with $\lambda_{\bf k}^2 > 0$.
Some simple manipulations allow us to recast eq.~(\ref{eq3:eigen_eq}) as follows
\begin{equation}
\left[-{d^2 \over d \theta^2}
-d \cot \theta {d \over d \theta}
+{1\over \sin^2\theta}\lambda_{\bf k}^2-m^{2} \right]\phi_{n,{\bf k}}(\theta) = -\alpha_{n,{\bf k}}^2 \phi_{n,{\bf k}}(\theta)~.\nonumber
        \label{eq5:leg_eq}
\end{equation}
By means of the following transformation,
\bea
\phi_{n,{\bf k}}(\theta)&=&
\sin^q\theta \psi_{n,{\bf k}}(\theta)~,
\eea
with $q=(1-d)/2$, the previous equation can be cast in standard Legendre form
\bea
{d^2\over d\theta^2}\psi_{n,{\bf k}}(\theta) +\cot\theta {d\over d\theta}\psi_{n,{\bf k}}(\theta)
+ \left[ \left( {d^2-1\over 4}+m^{2}-\alpha_{n,{\bf k}}^2 \right)
-\left(
\lambda_{\bf k}^2 + {(d-1)^2\over 4}
\right)
\csc^2 \theta
\right] \psi_{n,{\bf k}}(\theta)=0~.\nonumber
\eea
Introducing the quantities
\bea
\mu_{\bf k}&=&\sqrt{q^2 +\lambda_{\bf k}^2}~,
\\
\nu_{n,\bf k}&=&-{1\over 2}\pm i \sqrt{\alpha_{n,{\bf k}}^2-\sigma^2}\equiv -{1\over 2}\pm i\omega_{n,{\bf k}}~,\label{eq:8}
\eea
with $\sigma^{2}=m^{2}+d^{2}/4$,
and noticing that $\Re \left[\mu_{\bf k}\right]\geq 0$ and $\Re \left[\nu_{n,\bf k}\right] \geq -1/2$, two linearly independent solutions to the above equation can be written in terms of (see Theorem 12.1, ref.~\cite{olver})
\bea
(1-z^2)^{q/2} \mbox{P}_{-{1\over 2}+ i\omega_{n,{\bf k}}}^{- \mu_{\bf k}}\left(z\right)~,~~(1-z^2)^{q/2}\mbox{Q}_{-{1\over 2}+ i\omega_{n,{\bf k}}}^{+ \mu_{\bf k}}\left(z\right)~,
\eea
we have used the symmetry property $\mbox{P}^{\pm \mu}_{\nu}=\mbox{P}^{\pm \mu}_{-1-\nu}$ and defined $z=\cos\theta$. The two functions above correspond to the {\it Ferrers} representation of the Legendre function (see Chapt.~5, Sec.~15 of ref.~\cite{olver}) and satisfactorily cover the interval $-1 \leq z \leq 1$, with cuts from $(-\infty, -1]$ and $[1,+\infty)$. By looking at the $z\rightarrow 1$ behavior of the above functions, it is easy to see that the solution proportional to $\mbox Q$ diverges in this limit. Thus, requirement of regularity at $\theta=0$ restricts the general solution to be
\begin{equation}
\varphi_{n,{\bf k}}(z,X_i) =
(1-z^2)^{q/2} \mbox{P}_{-{1\over 2}+ i\omega_{n,{\bf k}}}^{- \mu_{\bf k}}\left(z\right)
{\cal H}_{{\bf k}}(X_i)~.
        \label{eq4:gensol}
\end{equation}
The solution (\ref{eq4:gensol}), aside for the requirement of regularity at $\theta=0$, is general. In the following discussion, for definiteness, we will consider the case of Dirichlet boundary conditions, thus imposing
on the eigenfunctions the relation
\bea
\mbox{F}_{\bf k}\left(\alpha_{n,\bf k}\right)\equiv \mbox{P}_{-1/2+i\omega_{n,\bf k}}^{-\mu_{\bf k}} \left(\cos \theta_0\right)=0~.
\label{eig:2-1}
\eea
Solutions to the previous equation implicitly determine the coefficients $\omega_{n,\bf k}$ and, via (\ref{eq:8}), the eigenvalues $\alpha_{n,\bf k}$. It should be easy to see that other boundary conditions can be easily be treated along the same lines as for Dirichlet.

\section{Zeta function on $\Omega$}
\label{sec3}
\begin{figure}[t]
	\begin{center}
	    \includegraphics[scale=1.25]{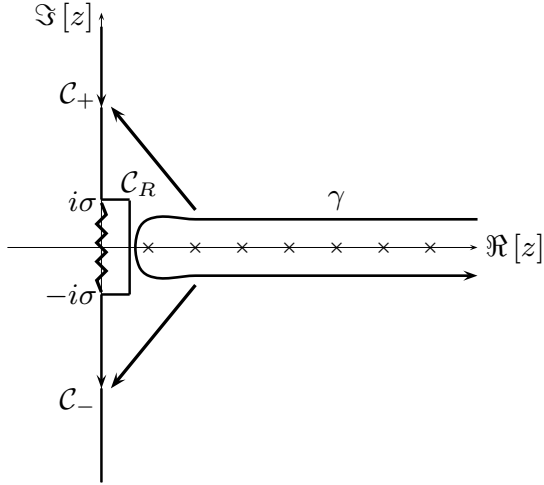}
	\end{center}
\caption{The figure illustrates the contour of integration $\gamma$ and the deformation used in the
calculation.}
\label{circuito}
\end{figure}

In this section we wish to present a formalism to compute the zeta function for the Laplacian on $\Omega$. The method we adopt is different from the one described, for example, in ref.~\cite{bordag}. The common step is the use of a contour integral representation for the spectral sum,
\bea
\zeta (s) = \sum_{n,{\bf k}} d({\bf k}) \alpha_{n,\bf k}^{-2s}=\sum_{n,{\bf k}} d({\bf k}) (\omega_{n,\bf k}+\sigma^{2})^{-s}~,
\label{zo}
\eea
but it will differ in the method of ref.~\cite{bordag} otherwise. The goal is to express the above zeta function in terms of the auxiliary zeta function defined on the base manifold ${\mathscr{N}}$ in the spirit of ref.~\cite{cheeger83},
\bea
\zeta_{\mathscr{N}} (s) = \sum_{{\bf k}} d({\bf k}) \mu_{\bf k}^{-2s}~.
\label{zn}\eea
The use of the residue theorem allows us to express the zeta function (\ref{zo}) as a contour integral:
\bea
\zeta (s) ={1\over 2\pi i} \sum_{\bf k} d({\bf k}) \oint_{\gamma} dz (z^2 +\sigma^2)^{-s} {\partial \over \partial z} \ln \mbox{F}_{\bf k}(z)~,\label{circgamma}
\eea
where the circuit $\gamma$ (see fig.~\ref{circuito}) encloses all the roots of (\ref{eig:2-1}). As it stands, the above integral representation is valid in the region $\Re\left[s\right]> (d+1)/2$.
By appropriately deforming the contour of integration along the imaginary axis and by utilizing the property $\mbox{P}^{\pm \mu}_{\nu}=\mbox{P}^{\pm \mu}_{-1-\nu}$, the above expression for the zeta function can be recast in the following form
\bea
\zeta (s) &=& {\sin \pi s \over \pi} \sum_{\bf k} d({\bf k})\int_\sigma^\infty dw (w^{2}-\sigma^{2})^{-s} {\partial \over \partial w}
\ln \mbox{P}_{-1/2+w}^{-\mu_{\bf k}} \left(\cos \theta_0\right)
\nonumber\\&&
+\sum_{\bf k} d({\bf k})\int_{C_{R}}{dz \over 2\pi i} (z^2 +\sigma^2)^{-s} {\partial \over \partial z} \ln \mbox{F}_{\bf k}(z)~,
\label{1}\eea
where the contour is deformed as illustrated in Fig.~\ref{circuito} and $C_{R}$ is a small rectangle enclosing the cut on the imaginary axis. The circuit is closed at infinity by a semi-circle, not shown in the figure, whose contribution to the contour integral vanishes due to the behavior of the integrand at large distance (the reader can check this behavior in the asymptotic expansion shown below). A simple calculation shows that the integral over ${\cal C}_R$ does not contribute to $\zeta(s)$. The first integral in the above expression comes, instead, from the portions $C_+$ and $C_-$ of the deformed circuit, and, by performing the change of variables
$\mu_{\bf{k}}^{2}u^{2}=w^{2}-\sigma^{2}$ in the expression (\ref{1}), one readily obtains
\bea
\zeta (s) = {\sin \pi s \over \pi} \sum_{\bf k} d({\bf k}) \mu_{\bf k}^{-2s} \int_0^\infty \frac{ du}{ u^{2s}} {\partial \over \partial u}
\ln \mbox{P}_{-1/2+\sqrt{u^2\mu_{\bf k}^2+\sigma^2}}^{-\mu_{\bf k}} \left(\cos \theta_0\right)~.
\label{zet}
\eea
We want to show, now, that the previous expression is convergent when the parameter $s$ is large enough. To this aim, let us use the large $\mu_k$ expansion of the generalized Legendre function \cite{flachi1,khusnutdinov03,thorne57}. Using a WKB form for the solution of the Legendre differential equation, one obtains a uniform asymptotic expansion for the Legendre functions of large order. Explicitly, one gets
\bea
\mbox{P}_{-1/2+\sqrt{u^2\mu_{\bf k}^2+\sigma^2}}^{-\mu_{\bf k}} \left(\cos \theta_0\right)\sim \sqrt{t\over 2\pi\mu_{\bf k}} e^{\mu_{\bf k} \tau}(\mu_{{\bf k}}u)^{-\mu_{\bf k}}\sum_{n=0}^{\infty} \frac{A_n(\nu)} { \mu_{\bf k}^{n}}\;,
\eea
where we have defined
\bea\label{willy6}
t\equiv t(u) &=& {1\over \sqrt{1+u^2\sin^2\theta_0}}~,\quad\nu=t\cos\theta_{0}\nonumber\\
\tau\equiv \tau(u) &=& 1+\ln \left({u\sin\theta_0 \over \sqrt{1+u^2\sin^2\theta_0} + \cos \theta_0}\right)
-u\left[ \arctan\left(\frac{1}{u}\right) - \arctan\left({1 \over u t \cos \theta_0}\right)
\right]~.
\eea
The polynomials $A_{n}(\nu)$ can be found to satisfy the following relation \cite{khusnutdinov03}
\bea
\sum_{n=0}^{\infty}\mu_{{\bf k}}^{-n}A_{n}(\nu)=\exp\left\{-\sum_{j=1}^{\infty}\frac{B_{2j}}{2j(2j+1)\mu_{{\bf k}}^{2j-1}}\right\}\sum_{j=0}^{\infty}\mu_{{\bf k}}^{-j}{\cal A}_{j}(\nu)\;,
\eea
where $B_{2j}$ are the Bernoulli numbers and the polynomials ${\cal A}_{n}(\nu)$ obey the recurrence relation
\bea\label{2}
{\cal A}_{n+1}(\nu)&=&\frac{(1-\nu^{2})(\nu^{2}u^{2}+1)}{2(u^{2}+1)}\partial_{\nu}\left[{{\cal A}}_{n}(\nu)\right]
-\frac{u^{2}}{8(u^{2}+1)}\int_{1}^{\nu}d\nu^{\prime}\left(5\nu'^{2}+\frac{1}{u^{2}}-1-\frac{4\sigma^{2}(1+u^{2})}{u^{2}(1+u^{2}\nu'^{2})}\right){\cal A}_{n}(\nu')\;,\nonumber
\eea
with ${\cal A}_{0}(\nu)=1$. For the interested reader, few of the first $A_{n}(\nu)$ are listed in the appendix of \cite{khusnutdinov03} and higher order ones can be obtained with a simple computer program.
It is, a this point, straightforward to obtain an asymptotic expansion for the logarithm of the Legendre function. From the previous results we have
\bea
\ln \mbox{P}_{-1/2+\sqrt{u^2\mu_{\bf k}^2+\sigma^2}}^{-\mu_{\bf k}} \left(\cos \theta_0\right)
\sim
\ln \left( \sqrt{t\over 2\pi\mu_{\bf k}} e^{\mu_{\bf k} \tau}(\mu_{\bf k}u)^{-\mu_{\bf k}}
\right) +
\sum_{n=1}^{\infty} {a_n(\nu) \over \mu_{\bf k}^{n}}~,
\label{exp}
\eea
where the polynomials $a_{n}(\nu)$ are defined through the cumulant expansion
\bea\label{3}
-\sum_{j=1}^{\infty}\frac{B_{2j}}{2j(2j-1)\mu_{{\bf k}}^{2j-1}}+\log\left\{1+\sum_{n=1}^{\infty}\mu_{{\bf k}}^{-n}{\cal A}_{n}(\nu)\right\}
\sim\sum_{k=1}^{\infty}\frac{a_{k}(\nu)}{\mu_{\bf k}^{k}}\;.
\eea
The previous expansion is uniform and valid for all ranges of variation of $u$ with the coefficients being bounded functions in $u$. Substituting the previous expression in formula (\ref{zet}) one obtains the following asymptotic formula for the zeta function:
\bea
\zeta (s) \sim {\sin \pi s \over \pi} \sum_{\bf k}\frac{ d({\bf k})}{ \mu_{\bf k}^{2s}} \int_0^\infty {du \over u^{2s}} {\partial \over \partial u}
\left(\frac{1}{2}\ln\frac{ t}{2\pi\mu_{{\bf k}}} - \ln(\mu_{\bf k} u)^{\mu_{\bf k}} + \mu_{\bf k} \tau + \sum_{n=1}^{\infty} {a_n(u)\over \mu_{\bf k}^{n}}\right)~.
\label{asymptzet}
\eea
Since the $\mu_{\bf k}$ are the eigenvalues of a second order partial differential operator on a smooth compact manifold, ${\mathscr{N}}$, they behave, asymptotically, according to the  Weyl's estimate, as \cite{gilkey95}
\bea
\mu_{\bf k}^{d/2}\sim\frac{2^{d-1}\pi^{d/2}d\,\Gamma(d/2)}{\textrm{Vol}_{\mathscr{N}}} |{\bf k}|\;.
\eea
This means that the spectral sum (\ref{zn}) converges for $\Re\left[s\right]> d/2$.
Therefore, by using the fact that the coefficients $a_n(u)$ are bounded functions, proving that the above expression is convergent, for $\Re\left[s\right]$ large enough, requires only trivial steps. In the convergency region for (\ref{asymptzet}), we can safely invert the summation and integration operation and write:
\bea
\zeta (s) = {\sin \pi s \over \pi} \int_0^\infty\frac{ du}{ u^{2s}} {\partial {\cal G}\over \partial u} (u,s)~.
\label{zetra}
\eea
where for notational convenience we have defined
\bea
{\cal G}(u,s)=
\sum_{\bf k} d({\bf k}) \mu_{\bf k}^{-2s}  \ln \mbox{P}_{-1/2+\sqrt{u^2\mu_{\bf k}^2+\sigma^2}}^{-\mu_{\bf k}} \left(\cos \theta_0\right)~.\label{intgrnd}
\eea
It should be clear, by looking at formula (\ref{asymptzet}), that in the asymptotic region defined by the expansion (\ref{exp}), the function ${\cal G}(u,s)$ can be expressed in terms of the base zeta function, $\zeta_{\mathscr{N}}(s)$. Further properties of ${\cal G}(u,s)$ can be obtained by using the Mittag-Leffler expansion for the base zeta function
\bea
\zeta_{\mathscr{N}}(s)={1\over \Gamma\left(s\right)}\sum_{p=0}^\infty{C_{p/2}\over s-d/2+p/2} + \varphi(s)~,\label{mle}
\eea
where $\varphi(s)$ is an entire function and $C_{p/2}$ are the integrated heat-kernel coefficients of the Laplacian $\nabla_\Sigma$ on the base manifold ${\mathscr{N}}$. Using the above expression, one cas see that
$\cal G$ will develop, at most, a pole in $s$ in the process of analytic continuation to $s\rightarrow 0$, {\it i.e.}:
\bea
{\cal G}(u,s) = {1 \over s}{\cal G}_{P}(u)+{\cal G}_{R}(u) + O(s)~,
\eea
where we indicated with ${\cal G}_{P}(u)$ and ${\cal G}_{R}(u)$ the polar and regular part of ${\cal G}(u,s)$ for small $s$.

At this point we need to perform the analytic continuation of the zeta function to $s=0$. For this purpose we will use the following result \cite{barvinsky}:\\

{\bf Lemma.} {\it Let $f(x)$ be a function defined for $x\geq \epsilon$ with $\epsilon >0$ and analytic at $x=\epsilon$. Assume that $f(x)$ has the following general asymptotic behavior for $x\rightarrow\infty$:
\bea
f(x)=\sum_{k=1}^{\rho_k<N}\left( f_k+\bar{ f}_k\ln x\right)x^{\rho_k}+[f]_{\rm log} \ln x+ [ f]_{\rm reg}+O(x^{-1})~,\qquad\rho_k>0~,
\label{phix}
\eea
where the subscripts ${\rm log}$ and ${\rm reg}$ refer to the solely logarithmic and regular
(non-singular) parts of $f(x)$ in the large $x$ limit. Then, there exists the analytic
continuation of the integral
\bea
\int_\epsilon^\infty {dx\over x^s} {d\over dx}  f(x)={[ f]_{\rm log}\over s}+
[ f]^{\infty}_{\epsilon}
+O(s)~,
\label{continue}
\eea
where $[ f]^{\infty}_{\epsilon} \equiv [ f]_{\rm reg}- f(\epsilon)$.}
\\

Taking the asymptotic expansion (\ref{exp}) and Taylor expanding all the terms for large $u$, one can immediately see that the behavior (\ref{phix}) is exactly reproduced. In other words, both functions ${\cal G}_{P}$ and ${\cal G}_{R}$ admit an expansion of the type (\ref{phix}) for large $u$.
Applying the above lemma to ${\cal G}_{P}$ and ${\cal G}_{R}$, we obtain an expression for $\zeta(s)$ and $\zeta'(s)$ in terms of ${\cal G}_{P}$ and
${\cal G}_{R}$. More precisely for $s\to 0$ one has
\bea
\zeta(s)&=&\frac{1}{s}\left[{\cal G}_{P}\right]_{log}+\left[{\cal G}_{R}\right]_{log}+\left(\left[{\cal G}_{P}\right]_{reg} -{\cal G}_{P}(0)\right)\label{zeta01} \nonumber\\
&+&s\left\{\left(\left[{\cal G}_{R}\right]_{reg} -{\cal G}_{R}(0)\right) - \int_0^\infty d u^{2} \ln u^{2} \frac{d}{du^{2}}{\cal G}_{P}(u^{2})\right\}+O(s^2)\;,
\eea
and also
\bea\label{zeta03}
\zeta'(s)=-\frac{1}{s^{2}}\left[{\cal G}_{P}\right]_{log}+\left\{\left(\left[{\cal G}_{R}\right]_{reg} -{\cal G}_{R}(0)\right) - \int_0^\infty d u^{2} \ln u^{2} \frac{d}{du^{2}}{\cal G}_{P}(u^{2})\right\}+O(s)\;.
\eea
It is well known \cite{bordag,bruning84,bruning87,kirsten01} that for singular problems the heat kernel asymptotic expansion
presents, in general, logarithmic terms. This non-standard behavior yields to the appearance of a pole in the spectral zeta function for $s=0$.
In our case, this singular term is proportional to $\left[{\cal G}_{P}\right]_{log}$.
As we will see in the next section, $\zeta(s)$ on ${\cal M}$ has a pole at $s=0$ provided that
$\zeta_{\mathscr{N}}(s)$ has a pole at $s=-1/2$. By recalling the formula (\ref{mle}) it is straightforward to see that the residue of
$\zeta_{\mathscr{N}}(s)$ at $s=-1/2$ is proportional to the heat kernel coefficient $C_{(d+1)/2}$. This coefficient vanishes identically only if
the base manifold ${\mathscr{N}}$ is even-dimensional and without boundary. In all other cases it is, in general, non zero.
In these singular situations the zeta function cannot be directly used, in the standard fashion, in order to compute the $\zeta$-regularized
functional determinant and some generalizations need to be used \cite{cognola06,cognola04,kirsten06}.

In what follows,
we will assume that the logarithmic term in the heat kernel asymptotic expansion does not appear, this will allow a
standard definition of the functional determinant \cite{bordag}. From a more formal point of
view, as said, this means that $\zeta(s)$ is regular at $s=0$, which is the case if $\zeta_{\mathscr{N}} (s)$ is regular at $s=-1/2$.
As we will show in the next section, this this means that $\left[{\cal G}_{P}\right]_{log}$ vanishes identically, since it is proportional
to the residue of $\zeta_{\mathscr{N}}$ at the point $s=-1/2$.

With the previous remarks in mind, from (\ref{zeta01}) and (\ref{zeta03}) one can easily find that
\bea
\zeta(0) &=& \left[{\cal G}_{R}\right]_{log} +\left(\left[{\cal G}_{P}\right]_{reg} -{\cal G}_{P}(0)\right)\;, \label{zeta0}\\
\zeta'(0) &=& \left(\left[{\cal G}_{R}\right]_{reg} -{\cal G}_{R}(0)\right) - \int_0^\infty d u^{2} \ln u^{2} \frac{d}{du^{2}}{\cal G}_{P}(u^{2})\label{zetaprime0}~.
\eea
The above formulas are general. In the following sections we will consider the case of Riemann caps (\ref{eq1:metric}) and compute the terms that are necessary for the evaluation of $\zeta(0)$ and $\zeta'(0)$.

\section{General Results}
\label{sec4}
The results of the previous section offer a way to compute the quantities $\zeta(0)$ and $\zeta'(0)$ and, in turn, the functional determinant. Specifically, the uniform asymptotic expansion (\ref{exp}) can be used to obtain the various terms in formulas (\ref{zeta0}) and (\ref{zetaprime0}) and, thus, a general expression for the determinant, as we will show in the following sections.

\subsection{Logarithmic contribution}
Let us begin by computing the logarithmic contribution. According to the above Lemma, we need to expand the function $\cal G$ for large values of $\rho=u\mu_{\bf k}$ \cite{barvinsky}. The coefficient of the logarithm will then provide the term $\left[{\cal G}\right]_{log}$. This coefficient is, in principle, composed of two terms $\left[{\cal G}_R\right]_{log}$ and $\left[{\cal G}_P\right]_{log}$.
The first step is, then, to expand (\ref{exp}) for large $\rho$. From its explicit expression one gets
\bea\label{willy10}
\ln \mbox{P}_{-1/2+\sqrt{\rho^2+\sigma^2}}^{-\mu_{\bf k}} \left(\cos \theta_0\right)
\sim-\frac{1}{4}(1+2\mu_{\bf k})\ln \rho^{2}-\frac{1}{2}\ln (2\pi\sin\theta_0)
+\rho\arctan(\tan\theta_0)+\mu_{\bf k}^{n}\sum_{n=1}^{\infty}\frac{{{\cal F}(\theta_{0})}}{\rho^{n}}+\sum_{n=1}^{\infty}\frac{a_{n}(\nu)}{\mu_{\bf k}^{k}}\;,
\eea
where ${\cal F}(\theta_{0})$ are continuous functions of the parameter $\theta_0$ which are computable from the expansion of $\tau(u)$ in (\ref{willy6}), and the terms $a_{n}(\nu)$ behave
like $u^{-m}$ with $m>1$ for large $u$.

It is trivial to see that only the first term of the previous expression gives rise to $\log\rho^{2}$ terms. The remaining sum over the eigenvalues is not difficult to compute according to the definition of the base zeta function, leading to the result
\bea
\left[{\cal G}\right]_{log}(s) = -{1\over 4}\lim_{s\rightarrow 0}\left[\zeta_{\mathscr{N}}(s)+2\zeta_{\mathscr{N}}\left(s-\frac{1}{2}\right)\right]\;.
\label{glog}\eea
It is clear, from this formula, that $\left[{\cal G}_P\right]_{log}$ does not vanish if $\zeta_{\mathscr{N}}$ is not regular at $s=-1/2$.
In fact, as mentioned before, by denoting with $\textrm{Res}$ the residue, one would obtain
\bea
\left[{\cal G}_P\right]_{log} &=& -\frac{1}{2}\textrm{Res}\,\zeta_{{\mathscr{N}}}\left(-\frac{1}{2}\right)=\textrm{Res}\,\zeta(0)\;.
\eea
which is a relation that has been shown to hold also for the generalized cone \cite{bordag}.

In order to deal with a well defined functional determinant, we will assume that $\textrm{Res}\,\zeta_{{\mathscr{N}}}(-1/2)=0$. In this way,
by taking the limit as $s \rightarrow 0$, $\zeta_{\mathscr{N}}(s-1/2)$ is finite and $\zeta_{\mathscr{N}}(s)$ is regular, as formula (\ref{mle}) indicates.
This allows us to obtain the result
\bea
\left[{\cal G}_R\right]_{log} &=& -{1\over 4}\left[\zeta_{\mathscr{N}}(0)+2\zeta_{\mathscr{N}}\left(-\frac{1}{2}\right)\right]\;,\label{willy20}\\
\left[{\cal G}_P\right]_{log} &=& 0\;.
\eea

\subsection{Polar contribution}
The next term we need for our analysis is the polar contribution to the spectral zeta-function, namely $\left[{\cal G}_{P}\right]$, which is computed from the expansion of the expression  (\ref{exp}) for large $\mu_{{\bf k}}$ \cite{barvinsky}. In this way, the sum over
$\mu_{\bf k}$ can be performed and expressed in terms of the zeta-function on the base manifold $\zeta_{\mathscr{N}}$. Since the analytic structure of
$\zeta_{\mathscr{N}}$ is known, it will not be difficult to extract the polar part $\left[{\cal G}_{P}\right]$.
It is straightforward to show that the expansion of (\ref{exp}) in inverse powers on $\mu_{{\bf k}}$ is
\bea\label{willy1}
\ln \mbox{P}_{-1/2+\sqrt{u^2\mu_{\bf k}^2+\sigma^2}}^{-\mu_{\bf k}} \left(\cos \theta_0\right)
&\sim&
\frac{1}{2}\ln\,t-\frac{1}{2}\ln 2\pi\mu_{\bf k}+\mu_{\bf k}\tau(u)-\mu_{{\bf k}}\ln\,\mu_{{\bf k}}
-
\mu_{{\bf k}}\ln u +
\sum_{n=0}^{\infty} {a_n(\nu) \over \mu_{\bf k}^{n}}~.
\eea
By performing the sum over $\mu_{\bf k}$ in the previous expansion, from the basic definition of the base zeta function, one readily obtains
\bea
\label{willy7}
\sum_{\bf k} d({\bf k}) \mu_{\bf k}^{-2s}  \ln \mbox{P}_{-1/2+\sqrt{u^2\mu_{\bf k}^2+\sigma^2}}^{-\mu_{\bf k}} \left(\cos \theta_0\right)
&\sim& \frac{1}{2}\zeta_{\mathscr{N}}(s)\left(\ln t-\ln 2\pi\right)
+\zeta_{\mathscr{N}}\left(s-\frac{1}{2}\right)\left(\tau(u)-\ln u\right)
 +\frac{1}{4}\zeta'_{\mathscr{N}}(s)\nonumber\\
&&
+\frac{1}{2}\zeta'_{\mathscr{N}}\left(s-\frac{1}{2}\right)+\sum_{n=1}^{\infty}a_{n}(\nu)\zeta_{\mathscr{N}}\left(s+\frac{n}{2}\right)\;,
\eea
where we have used the following relation
\bea
\lim_{\hat q\rightarrow 1} \left[{d\over dq}\sum \sum_{\bf k} d({\bf k}) \mu_{\bf k}^{-2s+1-q}\right]_{q=\hat q}
= - \sum_{\bf k} d({\bf k}) \mu_{\bf k}^{-2s} \ln \mu_{\bf k}=\frac{1}{2}\zeta'_{{\mathscr{N}}}(s)~.
\eea
From the expansion (\ref{willy7}) we only keep the terms that will contribute to the polar part.
It is not difficult to see that the only potential terms contributing to the polar part $\left[{\cal G}_{P}\right]$ come
from the infinite series in (\ref{willy7}). In fact, by exploiting the Mittag-Leffler expansion (\ref{mle}) one can
soon realize that $\zeta_{\mathscr{N}}$ possesses poles of the first order for $k<d$ at $s=d/2,(d-1)/2,\ldots,1/2$ with residues \cite{elizalde,kirsten01,leseduarte94}
\bea
\textrm{Res}\zeta_{\mathscr{N}}(s)\Big|_{s=\frac{k-d}{2}}=\frac{C_{k}}{\Gamma\left(\frac{k-d}{2}\right)}\;, \textrm{where}\quad n=0,1,2\ldots,d-1\;.
\eea
This remark allows us to write an explicit expression for the polar part, namely
\bea\label{willy8}
{\cal G}_{P}(u)=\sum_{n=1}^{d}a_{n}(\nu)\textrm{Res}\,\zeta_{\mathscr{N}}\left(\frac{n}{2}\right)\;.
\eea

This relation can be used in order to find ${\cal G}_{P}(0)$ which is needed for the computation of $\zeta(0)$ as
formula (\ref{zeta0}) shows. By taking the limit as $u\to 0$ one simply obtains
\bea\label{willy22}
{\cal G}_{P}(0)=\sum_{n=1}^{d}a_{n}(\sigma,\theta_0)\textrm{Res}\,\zeta_{\mathscr{N}}\left(\frac{n}{2}\right)\;,
\eea
where we have used the fact that, according to the relations in (\ref{willy6}), $\lim_{u\to 0}\nu(u)=\cos\theta_0$. The terms $a_{n}(\sigma,\theta_0)$ are
polynomials in $\sin(\theta_0/2)$ and $\sigma$, and have the following general form
\bea\label{willy14}
a_{n}(\sigma,\theta_0)={\cal C}_{n}+\sum_{l=1\atop k=0}^{n}c_{l,k}\sigma^{2k}\sin^{2l}\left(\frac{\theta_{0}}{2}\right)\;.
\eea
Some of the first few $a_{n}(\cos\theta_0)$ are listed in the appendix. The numerical coefficients
${\cal C}_{n}$ can be found by taking the limit as $\theta_{0}\to 0$ of the previous expression. By analyzing, in this limit,
the recurrence relation (\ref{2}) and the expansion (\ref{3}) one can easily realize that
\bea
{\cal C}_{n}=-\frac{B_{n+1}}{n(n+1)}=\frac{\zeta_{R}(-n)}{n}\;,
\eea
where $\zeta_{R}(n)$ is the Riemann zeta function and the last equality can be proved by taking the limit as $q\to 0$ of the
functional relation \cite{gradshtein07}
\bea
\zeta_{H}(-n,q)=-\frac{B_{n+1}(q)}{n+1}\;,
\eea
valid for $n\geq 0$, with $\zeta_{H}(n,q)$ being the Hurwitz zeta function and $B_{n}(q)$ being the Bernoulli polynomials.

\subsection{Regular contribution}
\label{section2}
The regular term $\left[{\cal G}\right]_{reg}$ is instead obtained by isolating the coefficient of the $\rho^{0}$ term in the large $\rho$ expansion that we have obtained in (\ref{willy10}). One can easily see that regular part has, then, the following form
\bea
\left[{\cal G}\right]_{reg} (s)= -{1\over 2}\ln \left( 2\pi \sin\theta_0\right) \zeta_{\mathscr{N}}(s)\;.
\label{gregaux}
\eea
From this expression we can isolate the polar and regular parts of $\left[{\cal G}\right]_{reg}(s) $ as $s\to 0$. In more details we obtain
\bea
\left[{\cal G}_{R}\right]_{reg}=-{1\over 2}\ln \left( 2\pi \sin\theta_0\right) \zeta_{\mathscr{N}}(0) \;,
\eea
and
\bea\label{willy21}
\left[{\cal G}_{P}\right]_{reg}=0\;.
\eea

The other term necessary for the evaluation of $\zeta'(0)$ is ${\cal G}_{R}(0)$ which has the form
\bea\label{4}
{\cal G}_{R}(0) =\textrm{ PF}\left\{ \lim_{s\to 0}\sum_{\bf k} d({\bf k}) \mu_{\bf k}^{-2s}
\ln \mbox{P}_{\sigma-1/2}^{-\mu_{\bf k}} \left(\cos \theta_0\right)\right\}~,
\eea
where $\textrm{ PF}$ denotes the {\it partie finie} of the expression.
It is clearly not possible to find a closed expression for the above summation. However, some progress can be made by expressing the
Legendre function in terms of a hypergeometric function as follows,
\bea
\mbox{P}_{\sigma-1/2}^{-\mu_{\bf k}} \left(\cos \theta_0\right)=\frac{1}{\Gamma(\mu_{\bf k}+1)}\tan^{\mu_{\bf k}}\left(\frac{\theta_0}{2}\right){_2}F_{1}\left(-\sigma+\frac{1}{2},\sigma+\frac{1}{2},\mu_{\bf k}+1; \sin^{2}\left(\frac{\theta_0}{2}\right)\right)\;.\nonumber\\
\eea
By substituting the previous expression in (\ref{4}), one obtains
\bea\label{willy11}
{\cal G}_{R}(0,s)&=&\frac{1}{2}\zeta'_{\mathscr{N}}(s)+\ln\tan\left(\frac{\theta_0}{2}\right)\zeta_{{\mathscr{N}}}\left(s-\frac{1}{2}\right)-\textrm{ PF} \sum_{\bf k} d({\bf k}) \mu_{\bf k}^{-2s}\ln\Gamma(\mu_{\bf k})\nonumber\\
&+&\textrm{ PF} \sum_{\bf k} d({\bf k}) \mu_{\bf k}^{-2s}\ln{_2}F_{1}\left(-\sigma+\frac{1}{2},\sigma+\frac{1}{2},\mu_{\bf k}+1; \sin^{2}\left(\frac{\theta_0}{2}\right)\right)\;.
\eea
At this point we exploit the integral representation of the function $\ln\Gamma(\mu)$ \cite{gradshtein07,olver}
\bea
\ln\Gamma(\mu_{\bf k})=\left(\mu_{\bf k}-\frac{1}{2}\right)\ln\mu_{\bf k}-\mu_{\bf k}+\frac{1}{2}\ln 2\pi+\int_{0}^{\infty}dt\left(\frac{1}{2}-\frac{1}{t}+\frac{1}{e^{t}-1}\right)\frac{e^{-t\mu_{\bf k}}}{t}\;,
\eea
to obtain
\bea
\sum_{\bf k} d({\bf k}) \mu_{\bf k}^{-2s}\ln\Gamma(\mu_{\bf k})&=&-\frac{1}{2}\zeta'_{{\mathscr{N}}}\left(s-\frac{1}{2}\right)+\frac{1}{4}\zeta'_{{\mathscr{N}}}\left(s\right)-\zeta_{{\mathscr{N}}}\left(s-\frac{1}{2}\right)
+\frac{1}{2}\zeta_{{\mathscr{N}}}\left(s\right)\ln 2\pi+\textrm{ PF}\, \Lambda(s)\;,
\eea
where we have introduced the function
\bea\label{willy13}
 \Lambda(s)=\sum_{\bf k} d({\bf k}) \mu_{\bf k}^{-2s}\int_{0}^{\infty}dt\left(\frac{1}{2}-\frac{1}{t}+\frac{1}{e^{t}-1}\right)\frac{e^{-t\mu_{\bf k}}}{t}\;.
\eea
In order to find the finite part of $\Lambda(s)$ one can simply subtract from its full expression (\ref{willy13}) the polar part $\Lambda_{\textrm{pole}}$
which has been evaluated in (\ref{willy14}). Since $\Lambda$ does not depend on the parameter $\theta_0$, we obviously need to subtract from (\ref{willy13})
only the terms in (\ref{willy14}) which are independent on $\theta_{0}$. By doing so we find the following finite expression at $s=0$.
\bea\label{willy23}
\textrm{PF}\Lambda(0)=\lim_{s\to 0}\left(\Psi(s)+\frac{1}{s}\sum_{n=1}^{d}\frac{\zeta_{R}(-n)}{n}\textrm{Res}\zeta_{{\mathscr{N}}}\left(\frac{n}{2}\right)\right)\;.
\eea
For a more explicit evaluation of (\ref{willy23}) it is convenient to recast it
 in the following form
\bea\label{willy32}
\textrm{ PF}\,\Lambda(0)=\sum_{k}d({\bf k})\int_{0}^{\infty}dt\left[\sum_{n=1}^{d}\frac{\zeta_{R}(-n)}{n!}t^{n}+\frac{1}{2}-\frac{1}{t}+\frac{1}{e^{t}-1}\right]\frac{e^{-t\mu_{\bf k}}}{t}
-\sum_{n=1}^{d}\frac{\zeta_{R}(-n)}{n}\textrm{PF}\,\zeta_{{\mathscr{N}}}\left(\frac{n}{2}\right)\;,
\eea
where we have used the relation
\bea
\frac{1}{s}\textrm{Res}\zeta_{{\mathscr{N}}}\left(\frac{n}{2}\right)=\zeta_{{\mathscr{N}}}\left(\frac{n}{2}\right)-\textrm{PF}\,\zeta_{{\mathscr{N}}}\left(\frac{n}{2}\right)\;,
\eea
and the integral representation of the inverse powers of $\mu_{{\bf k}}$ as
\bea
\mu_{\bf k}^{-n}=\frac{1}{\Gamma(n)}\int_{0}^{\infty}dt\,t^{n-1}e^{-t\mu_{\bf k}}\;.
\eea
In order to deal with the integral in (\ref{willy32}) we will follow the methods developed in \cite{bordag,bordag96} where a similar object has been studied in the
context of the $d$-dimensional ball. At this point, it is convenient to define the
following {\it square root} of the heat kernel \cite{bordag96,fucci}
\bea
K_{\mathscr{N}}^{1/2}(t)=\sum d(\mu_{\bf k})e^{-t\mu_{\bf k}}\;.
\eea
By recalling that the spectral $\zeta$-function is obtained from the heat kernel by inverse Mellin transform as \cite{esposito97,kirsten01,vassile03}
\bea
\zeta_{\mathscr{N}}\left(\frac{s}{2}\right)=\frac{1}{\Gamma(s)}\int_{0}^{\infty}dt\,t^{s-1}K_{\mathscr{N}}^{1/2}(t)\;,
\eea
we can write the expression in (\ref{willy32}) as follows
\bea
\textrm{PF}\,\Lambda(0,z)&=&\sum_{n=1}^{d}\frac{\zeta_{R}(-n)}{n\Gamma(n)}\Gamma(z+n)\zeta_{\mathscr{N}}\left(\frac{z+n}{2}\right)+\frac{1}{2}\Gamma(z)\zeta_{\mathscr{N}}\left(\frac{z}{2}\right)
-\Gamma(z-1)\zeta_{\mathscr{N}}\left(\frac{z-1}{2}\right)\nonumber\\
&+&\Gamma(z)\zeta_{{\mathscr{N}}+1}(z)-\sum_{n=1}^{d}\frac{\zeta_{R}(-n)}{n}\textrm{PF}\,\zeta_{{\mathscr{N}}}\left(\frac{n}{2}\right)\;,
\eea
where we have introduced a regularizing parameter $z$ such that $\textrm{PF}\,\Lambda(0,z)\to \textrm{PF}\,\Lambda(0)$ as $z\to 0$,
and we have introduced a new function defined as \cite{bordag96,fucci}
\bea
\zeta_{\mathcal{N}+1}(z)=\frac{1}{\Gamma(z)}\int_{0}^{\infty}t^{z-1}\frac{K_{\mathcal{N}}^{1/2}(t)}{e^{t}-1}\;.
\eea
The limit of the above expression as $z\to 0$ can now be taken leading to the result
\bea\label{willy29}
\textrm{PF}\,\Lambda(0)&=&\sum_{n=1}^{d}\frac{\zeta_{R}(-n)}{n}\left[2\Psi(n)\textrm{Res}\,\zeta_{\mathscr{N}}\left(\frac{n}{2}\right)+\textrm{PF}\,\zeta_{\mathscr{N}}\left(\frac{n}{2}\right)\right]+\frac{1}{4}\zeta'_{\mathscr{N}}\left(0\right)+\frac{1}{2}\zeta'_{\mathscr{N}}\left(-\frac{1}{2}\right)+\zeta_{\mathscr{N}}\left(-\frac{1}{2}\right)\nonumber\\
&-&\frac{\gamma}{2}\zeta_{\mathscr{N}}\left(0\right)-\gamma\zeta_{\mathscr{N}}\left(-\frac{1}{2}\right)+\lim_{z\to 0}\frac{1}{z}\Bigg[2\sum_{n=1}^{d}\frac{\zeta_{R}(-n)}{n}\textrm{Res}\,\zeta_{\mathscr{N}}\left(\frac{n}{2}\right)+\frac{1}{2}\zeta_{\mathscr{N}}\left(0\right)\nonumber\\
&+&\zeta_{\mathscr{N}}\left(-\frac{1}{2}\right)+\Gamma(z)\zeta_{\mathcal{N}+1}(z)\Bigg]-\sum_{n=1}^{d}\frac{\zeta_{R}(-n)}{n}\textrm{PF}\,\zeta_{{\mathscr{N}}}\left(\frac{n}{2}\right)\;,
\eea
where $\Psi(n)$ is defined as the logarithmic derivative of the gamma function.
At this point, it is convenient to utilize the
series \cite{bordag96,gradshtein07}
\bea
\frac{1}{e^{t}-1}=\frac{1}{t}-\frac{1}{2}-\sum_{n=1}^{\infty}\frac{t^{n}}{n!}\zeta_{R}(-n)\;,
\eea
and the asymptotic expansion of the heat kernel $K_{\mathscr{N}}^{1/2}(t)$ for small values of $t$ as
\bea
K_{{\mathscr{N}}}^{1/2}(t)\sim\sum\mathscr{A}_{k}t^{k-d}\;,
\eea
where
\bea
\mathscr{A}_{k}=2\Gamma(d-k)\textrm{Res}\,\zeta_{{\mathscr{N}}}\left(\frac{d-k}{2}\right)\;,
\eea
for $k=0,1,\cdots ,d-1$, $k=d-(2l+1)$ with $l\in\mathbb{N}^{+}$, and
\bea
\mathscr{A}_{k}=\frac{(-1)^{k-d}}{(k-d)!}\zeta_{{\mathscr{N}}}\left(\frac{d-k}{2}\right)\;,
\eea
for $k>d$ and $k\in\mathbb{N}^{+}$. With the help of the above results, it is not very difficult to prove that \cite{bordag96}
\bea
\zeta_{{\mathscr{N}}+1}(0)=-\zeta_{\mathscr{N}}\left(-\frac{1}{2}\right)-\frac{1}{2}\zeta_{\mathscr{N}}(0)-2\sum_{n=1}^{d}\frac{\zeta_{R}(-n)}{n}\textrm{Res}\,\zeta_{\mathscr{N}}\left(\frac{n}{2}\right).
\eea
The last expression ensures that the limit as $z\to 0$ in (\ref{willy29}) vanishes identically, as expected, yielding the final result
\bea\label{willy100}
\textrm{PF}\,\Lambda(0)&=&\zeta'_{\mathcal{N}+1}(0)+\frac{1}{4}\zeta'_{\mathcal{N}}(0)+\frac{1}{2}\zeta'_{\mathcal{N}}\left(-\frac{1}{2}\right)+\zeta_{\mathcal{N}}\left(-\frac{1}{2}\right)
+2\sum_{n=1}^{d}\frac{\zeta_{R}(-n)}{n}\left[\sum_{k=1}^{n-1}\frac{1}{k}\right]\textrm{Res}\,\zeta_{\mathscr{N}}\left(\frac{n}{2}\right)\;,
\eea
where in the formula (\ref{willy100}) we have used the fact that \cite{gradshtein07}
\bea
\Psi(n)=-\gamma+\sum_{k=1}^{n-1}\frac{1}{k}\;.
\eea

In order to deal, now, with the last term in expression (\ref{willy11}) we use an expedite method relying on the Abel-Plana summation formula \cite{barvinsky,flachi1}.
Let us define the function
\bea\label{willy24}
\varphi({\bf x})\equiv d({\bf x})
\ln {_2}F_{1}\left(-\sigma+\frac{1}{2},\sigma+\frac{1}{2},\mu_{\bf x}+1; \sin^{2}\left(\frac{\theta_0}{2}\right)\right)\;.
\eea
From the series representation of the hypergeometric function and from the fact that
$d(\bf x)$ is polynomial in ${\bf x}$, one can see that our function $\gamma({\bf x})$ uniformly satisfies the condition $\lim_{y\to 0}e^{-2\pi |y|}|\gamma(x+iy)|=0$. This allows us to apply the Abel-Plana formula as follows \cite{barvinsky,olver}
\bea\label{willy12}
\lefteqn{
\textrm{ PF} \sum_{\bf k} d({\bf k}) \mu_{\bf k}^{-2s}\ln{_2}F_{1}\left(-\sigma+\frac{1}{2},\sigma+\frac{1}{2},\mu_{\bf k}+1; \sin^{2}\left(\frac{\theta_0}{2}\right)\right)
}
\nonumber\\&&
=\textrm{ PF} \int_0^\infty\frac{\varphi(x)}{\mu^{2s}_{\bf x}}dx+i\int_{0}^{\infty}\frac{\varphi(ix)+\varphi(-ix)}{e^{2\pi x}-1}dx+\frac{1}{2}\varphi(0)\;,
\eea
where he have set $s=0$ in the last two terms because they are finite for $s\to 0$. The finite part of the first integral in (\ref{willy12})
can be obtained in the same way that was described for the function $\Psi(s)$ in (\ref{willy13}). By subtracting from the $s$-dependent integral in (\ref{willy12})
only the polar contributions in (\ref{willy14}) depending on $\theta_0$ one obtains
\bea\label{willy17}
\textrm{ PF} \int_0^\infty\frac{\varphi(x)}{\mu^{2s}_{\bf x}}dx= \int_0^\infty\frac{\varphi(x)}{\mu^{2s}_{\bf x}}dx-\frac{1}{s}\sum_{n=1}^{d}\textrm{Res}\zeta_{{\mathscr{N}}}\left(\frac{n}{2}\right)\sum_{l=1\atop k=0}^{n}c_{l,k}\sigma^{2k}\sin^{2l}\left(\frac{\theta_{0}}{2}\right)\;.
\eea
At this point, the above integral can be analyzed further, by utilizing repeated integration by parts,
once the base manifold ${\mathscr{N}}$ is specified and the degeneracy of the hyperspherical harmonics on ${\mathscr{N}}$ is known (see e.g. \cite{barvinsky92,flachi1}).

By collecting the relevant results obtained above, we can finally write an explicit expression for ${\cal G}_{R}(0)$ as follows
\bea
{\cal G}_{R}(0)&=&\zeta_{{\mathscr{N}}}\left(-\frac{1}{2}\right)\ln\tan\left(\frac{\theta_0}{2}\right)-\frac{1}{2}\zeta_{{\mathscr{N}}}\left(0\right)\ln 2\pi-\zeta'_{{\mathscr{N}}+1}\left(0\right)-2\sum_{n=1}^{d}\frac{\zeta_{R}(-n)}{n}\left[\sum_{k=1}^{n-1}\frac{1}{k}\right]\textrm{Res}\,\zeta_{\mathscr{N}}\left(\frac{n}{2}\right)\nonumber\\
&+&\textrm{ PF} \int_0^\infty\frac{\varphi(x)}{\mu^{2s}_{\bf x}}dx+i\int_{0}^{\infty}\frac{\varphi(ix)+\varphi(-ix)}{e^{2\pi x}-1}dx+\frac{1}{2}\varphi(0)\;.
\eea

The last term that is necessary in order to evaluate $\zeta'(0)$ is the integral in (\ref{zetaprime0}), which represents the non-local
contribution to $\zeta'(0)$.
By recalling the polar terms already found in (\ref{willy8}), we can write that
\bea
\int_0^\infty d u^{2} \ln u^{2} \frac{d}{du^{2}}{\cal G}_{P}(u^{2})=\sum_{n=1}^{d}\textrm{Res}\zeta_{{\mathscr{N}}}\left(\frac{n}{2}\right)\int_{0}^{\infty}du^{2} \ln u^{2} \frac{d}{du^{2}} a_{n}(\nu(u^{2}))\;.
\eea
The functions $a_{n}(\nu(u))$ can be computed, as many as needed, from the cumulant expansion (\ref{3}).

\subsection{$\zeta(0)$ and $\zeta'(0)$}
We are now able to write down an expression for $\zeta(0)$ and $\zeta'(0)$ by using the results obtained in the
previous sections. By recalling the expression (\ref{zeta0}) and the results (\ref{willy20}), (\ref{willy22}) and (\ref{willy21})
we have the following formula for the value of the zeta function on $\Omega$ at $s=0$
\bea\label{willy18}
\zeta(0)=-\frac{1}{4}\zeta_{{\mathscr{N}}}(0)-\frac{1}{2}\zeta_{{\mathscr{N}}}\left(-\frac{1}{2}\right)-\sum_{n=1}^{d}a_{n}(\sigma,\theta_0)\textrm{Res}\,\zeta_{\mathscr{N}}\left(\frac{n}{2}\right)\;,
\eea
where the polynomials $a_{n}(\sigma,\theta_0)$ are given in (\ref{willy14}).
In order to write an expression for the derivative of $\zeta(s)$ at $s=0$, we need the formula (\ref{zetaprime0}) and
the results obtained in Sec. \ref{section2}. The final expression is slightly more cumbersome and reads
\bea\label{willy19}
\zeta'(0)&=&-\left[\frac{1}{2}\zeta_{\mathscr{N}}(0)+\zeta_{\mathscr{N}}\left(-\frac{1}{2}\right)\right]\ln(\sin\theta_0)+\zeta_{\mathscr{N}}\left(-\frac{1}{2}\right)\ln(1+\cos\theta_0)+\zeta'_{{\mathscr{N}}+1}(0)\nonumber\\
&+&2\sum_{n=1}^{d}\frac{\zeta_{R}(-n)}{n}\left[\sum_{k=1}^{n-1}\frac{1}{k}\right]\textrm{Res}\,\zeta_{\mathscr{N}}\left(\frac{n}{2}\right)-\sum_{n=1}^{d}\textrm{Res}\,\zeta_{{\mathscr{N}}}\left(\frac{n}{2}\right)\int_{0}^{\infty}du^{2} \ln u^{2} \frac{d}{du^{2}} a_{n}(\nu(u^{2}))
\nonumber\\
&-&\frac{1}{2}\varphi(0)-i\int_{0}^{\infty}\frac{\varphi(ix)+\varphi(-ix)}{e^{2\pi x}-1}dx-\lim_{s\to 0}\left\{\int_0^\infty\frac{\varphi(x)}{\mu^{2s}_{\bf x}}dx-\frac{1}{s}\sum_{n=1}^{d}\textrm{Res}\zeta_{{\mathscr{N}}}\left(\frac{n}{2}\right)\sum_{l=1\atop k=0}^{n}c_{l,k}\sigma^{2k}\sin^{2l}\left(\frac{\theta_{0}}{2}\right)\right\}\;,
\eea
where the function $\varphi(x)$ has been defined in (\ref{willy24}) and the expression in the last line is finite, by construction, as $s\to 0$.

Although the above expressions are quite implicit, they are very general and valid for arbitrary dimension and for any smooth and compact base manifold ${\mathscr{N}}$. We would like to stress, here, that more explicit formulas can be obtained once the base manifold ${\mathscr{N}}$ is specified.
In the next section we will consider a particular case in which the base manifold is a $d$-dimensional ball.
This case is of special interest because the spectral zeta function on $d$-dimensional ball can be explicitly
evaluated in terms of the Barnes zeta function \cite{bordag,chang}.

\section{A particular case: $d$-dimensional ball as base manifold}
\label{sec5}
In this section we will consider the case in which the base manifold is a $d$-dimensional ball. In this particular
situation the eigenvalues of the Laplacian on ${\mathscr{N}}$ are known to be
\bea
\mu_{{\bf k}}=\left(k+\frac{d-1}{2}\right)\;,
\label{zin}
\eea
with $k\geq 0$, and the eigenfunctions are hyperspherical harmonics with degeneracy
\bea\label{willy300}
d({\bf k})=(2k+d-1)\frac{(k+d-2)!}{k!(d-1)!}\;.
\label{zan}
\eea
The explicit knowledge of the eigenvalues $\mu_{{\bf k}}$ and their degeneracy allows us
to write, by using the definition (\ref{zn}), the zeta function on ${\mathscr{N}}$ as follows
\bea
\zeta_{{\mathscr{N}}}(s)=\sum_{k=0}^{\infty}(2k+d-1)\frac{(k+d-2)!}{k!(d-1)!}\left(k+\frac{d-1}{2}\right)^{-2s}\;.
\eea

It is straightforward to show, with the use of algebraic manipulations on the factorials, that the above zeta function
can be written in terms of a sum of Barnes zeta functions \cite{bordag,chang}
\bea\label{willy30}
\zeta_{{\mathscr{N}}} (s) =\zeta_{\mathcal{B}}\left(2s,\frac{d+1}{2}\right)+\zeta_{\mathcal{B}}\left(2s,\frac{d-1}{2}\right)\;,
\eea
where $\zeta_{\mathcal{B}}(s,a)$ is defined as \cite{barnes,dowker05}
\bea
\zeta_{{\cal B}}(s,a|\vec{r})=\sum_{\vec{m}=0}^{\infty}\frac{1}{(a+\vec{m}\cdot\vec{r})^{s}}\;,
\eea
valid for $\Re(s)>d$ where $\vec{m}$ and $\vec{r}$ are $d$-dimensional vectors, and where the notation $\zeta_{{\cal B}} (s,a|\vec 1)=\zeta_{{\cal B}} (s,a)$ has been used.
In order to extract the information we need from the Barnes zeta function we will utilize its integral representation as follows \cite{bordag,fucci}
\bea
\zeta_{{\cal B}}(s,a)=\frac{i\Gamma(1-s)}{2\pi}\int_{L}dy\,\frac{e^{y\left(\frac{d}{2}-a\right)}(-y)^{s-1}}{2^{d}\sinh^{d}\left(\frac{y}{2}\right)}\;,
\eea
where $L$ represents the Hankel contour. With the help of the above representation and the relation (\ref{willy30}) we
can write an expression for $\zeta_{{\mathscr{N}}}(s)$ as
\bea
\zeta_{{\mathscr{N}}}(s)=\frac{i\Gamma(1-2s)}{2\pi}\int_{L}dy\,\frac{(-y)^{2s-1}\cosh\left(\frac{y}{2}\right)}{2^{d-1}\sinh^{d}\left(\frac{y}{2}\right)}\;.
\eea
At this point it is suitable to make a change of variables, $y/2\to y$, to recast the previous formula
in the form \cite{bordag,fucci}
\bea\label{willy31}
\zeta_{{\mathscr{N}}}(s)=(-1)^{2s-2}\frac{i\Gamma(2-2s)}{2\pi(d-1)}2^{2s+1-d}\sum_{\nu=0}^{\infty}\frac{D_{\nu}^{(d-1)}}{\nu!}\int_{L}dy\,y^{2s-d-1+\nu}\;,
\eea
where the coefficients $D_{\nu}^{(d-1)}$ can be easily computed, by equating like powers of $y$, from the formula
\bea
\left(\frac{y}{\sinh y}\right)^{d-1}=\sum_{\nu=0}^{\infty}D_{\nu}^{(d-1)}\frac{y^{\nu}}{\nu!}\;.
\eea

The integral representation (\ref{willy31}) is particularly useful for computing the residue of $\zeta_{{\mathscr{N}}}$
at different values of $s$. Specifically, we need the residue of $\zeta_{{\mathscr{N}}}(s)$ at the points $s=m/2$ with $m$ being a
positive integer. It is not very difficult to see that the integral in (\ref{willy31}) vanishes unless $\nu=d-m$ where the integrand
has a simple pole. By simply using the residue theorem we get \cite{fucci}
\bea
\textrm{Res}\,\zeta_{{\mathscr{N}}}\left(\frac{m}{2}\right)=\frac{2^{m-d}D_{d-m}^{(d-1)}}{(d-1)(m-2)!(d-m)!}\;,
\eea
valid for $m\geq 2$ and $d\geq m$. Moreover, we will need the values of $\zeta_{{\mathscr{N}}}$ at the points $s=0$ and
$s=-1/2$. From the expression (\ref{willy31}) one easily gets
\bea
\zeta_{\mathscr{N}}(0)=-\frac{2^{1-d}D^{(d-1)}_{d}}{(d-1)d!}\;,
\eea
and
\bea
\zeta_{{\mathscr{N}}}\left(-\frac{1}{2}\right)=\frac{2^{1-d}D_{d+1}^{(d-1)}}{(d-1)(d+1)!}\;.
\eea
These last formulas immediately give an expression for $\zeta(0)$ for arbitrary dimension $d$. More precisely,
by recalling (\ref{willy14}) and (\ref{willy18}), one obtains
\bea
\zeta(0)&=&\frac{1}{2^{d+1}(d-1)d!}\left[D_{d}^{(d-1)}-\frac{2}{d+1}D_{d+1}^{(d-1)}\right]-\frac{1}{2^{d}(d-1)}\sum_{n=2}^{d}\frac{2^{n}\zeta_{R}(-n)D_{d-n}^{(d-1)}}{n(n-2)!(d-n)!}\nonumber\\
&-&\frac{1}{2^{d}(d-1)}\sum_{n=2}^{d}\frac{2^{n}D_{d-n}^{(d-1)}}{(n-2)!(d-n)!}\sum_{l=1\atop k=0}^{n}c_{l,k}\sigma^{2k}\sin^{2l}\left(\frac{\theta_{0}}{2}\right)\;.\nonumber\\
\eea

Let us, now, turn our attention to the term $\zeta_{{\mathscr{N}}+1}(s)$ that appears in the general result (\ref{willy19}).
In the case of a $d$-dimensional ball as base manifold we have that \cite{bordag}
\begin{equation}
\zeta_{\mathscr{N}+1}(s)=\sum_{k=0}^{\infty}e(k)\left(k+\frac{d+1}{2}\right)^{-s}\;,
\end{equation}
where we have introduced the coefficients
\begin{equation}
e(k)=(2k+d)\frac{(k+d-1)!}{k!d!}.\nonumber
\end{equation}
By writing \cite{bordag96,fucci}
\begin{equation}\label{52}
e(k)= \sum_{\alpha=0}^{d}e_{\alpha}(d)\left(k+\frac{d+1}{2}\right)^{\alpha}\;,
\end{equation}
which defines $e_\alpha(d)$, we have that
\begin{equation}
\zeta_{\mathscr{N}+1}(s)=\sum_{\alpha=0}^{d}e_{\alpha}(d)\zeta_{H}\left(s-\alpha,\frac{d+1}{2}\right)\;,
\end{equation}
where the coefficients $e_{\alpha}(d)$ depend on the dimension $d$ and are determined from the equation (\ref{52}), and $\zeta_{H}$ represents the
Hurwitz $\zeta$-function.
In particular, for its derivative at $s=0$ we obtain \cite{bordag96,fucci}
\begin{equation}
\zeta^{\prime}_{\mathscr{N}+1}(0)=\sum_{\alpha=0}^{d}e_{\alpha}(d)\zeta^{\prime}_{H}\left(-\alpha,\frac{d+1}{2}\right)\;.
\end{equation}
This expression, involving the derivatives of the Hurwitz zeta function, can be rewritten in terms of Riemann zeta function and its first derivative.
In order to do so, we must distinguish between even and odd dimensional cases. For even dimensions, namely when $d=2q$ with $q\in\mathbb{N}^{+}$, we get \cite{fucci}
\bea\label{sam}
\zeta'_{H}\left(-\alpha,\frac{d+1}{2}\right)=\zeta'_{H}\left(-\alpha,q+\frac{1}{2}\right)&=&\frac{\ln\,2}{2^{\alpha}}\left[\zeta_{R}(-\alpha)-\sum_{n=1}^{2q-1}n^{\alpha}\right]+
(2^{-\alpha}-1)\zeta'_{R}(-\alpha)\nonumber\\
&&+\frac{1}{2^{\alpha}}\sum_{n=1}^{2q-1}n^{\alpha}\ln\,n-\sum_{n=1}^{q-1}n^{\alpha}\ln\,n\;,
\eea
while for odd dimensions, i.e. when $d=2q+1$ with $q\in\mathbb{N}^{+}$, we obtain
\bea\label{sam1}
\zeta'_{H}\left(-\alpha,\frac{d+1}{2}\right)=\zeta'_{H}\left(-\alpha,q+1\right)=\zeta'_{R}(-\alpha)+\sum_{n=1}^{q-1}(n+1)^{\alpha}\ln\,(n+1)\;.
\eea
We would like to mention that the derivatives of the Riemann zeta function appearing in (\ref{sam}) and (\ref{sam1}) can be further simplified
by differentiating the reflection formula
\bea
\zeta_{R}(1-s)=2(2\pi)^{-s}\Gamma(s)\cos\left(\frac{\pi s}{2}\right)\zeta_{R}(s)\;.
\eea
In fact, for $\alpha=2p$ where $p\in\mathbb{N}^{+}$, it is not very difficult to get \cite{miller,fucci09}
\bea
\zeta'_{R}(-2p)=(-1)^{p}\pi(2\pi)^{-2p-1}\Gamma(2p+1)\zeta_{R}(2p+1)\;,
\eea
and for $\alpha=2p+1$ where $p\in\mathbb{N}^{+}$, one obtains
\bea
\zeta'_{R}(-2p+1)=\frac{B_{2p}}{2p}\left[\Psi(2p)-\ln\,2\pi\right]+\frac{(-1)^{p+1}2(2p-1)!}{(2\pi)^{2p}}\zeta_{R}'(2p)\;.
\eea

The next term that can be more explicitly evaluated is the integral in (\ref{willy17}). By recalling the expression for the degeneracy of the eigenvalues (\ref{willy300}) we can rewrite the function (\ref{willy24}) as follows
\bea
\varphi(x)=x^{d}\sum_{\alpha=1}^{d-1}e_{\alpha}(d)x^{\alpha-d}\left(1+\frac{d-1}{2x}\right)^{\alpha}
\ln {_2}F_{1}\left(-\sigma+\frac{1}{2},\sigma+\frac{1}{2},\mu_{\bf x}+1; \sin^{2}\left(\frac{\theta_0}{2}\right)\right)\;.
\eea
This formula allows us to write
\bea\label{willy35}
\int_0^\infty\frac{\varphi(x)}{\mu^{2s}_{\bf x}}dx=\int_{0}^{\frac{2}{d-1}}dx\,x^{2s-d-2}\phi(x)\;,
\eea
where we have performed a change of variables to $1/x$ and we have defined
\bea
\phi(x)=\sum_{\alpha=1}^{d-1}e_{\alpha}(d)x^{d-\alpha}\left(1+\frac{d-1}{2}x\right)^{\alpha}
\ln {_2}F_{1}\left(-\sigma+\frac{1}{2},\sigma+\frac{1}{2},\frac{1}{x}+1; \sin^{2}\left(\frac{\theta_0}{2}\right)\right)\;,
\eea
which is an analytic function at $x=0$. It is convenient, at this point, to split the integral (\ref{willy35}) into three pieces \cite{barvinsky,flachi1}
as
\bea
\int_0^\infty\frac{\gamma(x)}{\mu^{2s}_{\bf x}}dx&=&\int_{0}^{1}dx\,x^{2s-d-2}\phi(x)+\int_{1}^{\infty}dx\,x^{-d-2}\phi(x)
-\int_{\frac{2}{d-1}}^{\infty}dx\,x^{-d-2}\phi(x)\;,
\eea
where we have set $s=0$ in the last two integrals because they yield a finite result in that limit. The first two terms can be integrated
by parts $d+2$ times to get \cite{barvinsky,flachi1}
\bea
\int_0^\infty\frac{\gamma(x)}{\mu^{2s}_{\bf x}}dx&=&\Bigg[\frac{x^{2s-d-1}}{2s-d-1}\phi(x)-\frac{x^{2s-d}}{(2s-d-1)(2s-d)}\frac{d\phi(x)}{dx}\nonumber\\
&+&\cdots-(-1)^{d}\frac{x^{2s}}{(2s-d-1)(2s-d)\cdots 2s}\frac{d^{d+1}\phi(x)}{dx^{d+1}}\Bigg]_{0}^{1}\nonumber\\
&+&(-1)^{d}\frac{1}{(2s-d-1)(2s-d)\cdots 2s}\int_{0}^{1}dx\,x^{2s}\frac{d^{d+2}\phi(x)}{dx^{d+2}}\nonumber\\
&+&\Bigg[\frac{x^{-d-1}}{(-d-1)}\phi(x)-\frac{x^{-d}}{(-d-1)(-d)}\frac{d\phi(x)}{dx}\nonumber\\
&+&\cdots-(-1)^{d}\frac{\ln x}{(-d-1)(-d)\cdots (-1)}\frac{d^{d+1}\phi(x)}{dx^{d+1}}\Bigg]_{1}^{\infty}\nonumber\\
&+&\frac{(-1)^{d}}{(-d-1)(-d)\cdots (-1)}\int_{1}^{\infty}dx\ln x\frac{d^{d+2}\phi(x)}{dx^{d+2}}-\int_{\frac{2}{d-1}}^{\infty}dx\,x^{-d-2}\phi(x)\;.
\eea
We can now perform the analytic continuation to $s=0$ of the expression above to get the compact result
\bea
\int_0^\infty\frac{\gamma(x)}{\mu^{2s}_{\bf x}}dx&=&\frac{1}{2s(d+1)!}\frac{d^{d+1}\phi(x)}{dx^{d+1}}\Bigg|_{x=0}+\frac{1}{(d+1)!}\left(\sum_{k=0}^{d}\frac{1}{d-k+1}\right)\frac{d^{d+1}\phi(x)}{dx^{d+1}}\Bigg|_{x=0}\nonumber\\
&-&\frac{1}{(d+1)!}\int_{0}^{\infty}dx\ln x\frac{d^{d+2}\phi(x)}{dx^{d+2}}-\int_{\frac{2}{d-1}}^{\infty}dx\,x^{-d-2}\phi(x)\;,
\eea
where the dependence on the intermediate cutoff $x=1$ has disappeared as expected. From this last formula, by recalling (\ref{willy17}), one can easily
find the relations
\bea
\sum_{n=1}^{d}\textrm{Res}\zeta_{{\mathscr{N}}}\left(\frac{n}{2}\right)\sum_{l=1\atop k=0}^{n}c_{l,k}\sigma^{2k}\sin^{2l}\left(\frac{\theta_{0}}{2}\right)=\frac{1}{2(d+1)!}\frac{d^{d+1}\phi(x)}{dx^{d+1}}\Bigg|_{x=0}\;,
\eea
and
\bea
\textrm{ PF} \int_0^\infty\frac{\varphi(x)}{\mu^{2s}_{\bf x}}dx&=&\frac{1}{(d+1)!}\left(\sum_{k=0}^{d}\frac{1}{d-k+1}\right)\frac{d^{d+1}\phi(x)}{dx^{d+1}}\Bigg|_{x=0}
-\frac{1}{(d+1)!}\int_{0}^{\infty}dx\ln x\frac{d^{d+2}\phi(x)}{dx^{d+2}}
\nonumber\\&&-
\int_{\frac{2}{d-1}}^{\infty}dx\,x^{-d-2}\phi(x)\;.
\eea

\subsection{Specific dimensions}

In this section we will list the results for $\zeta(0)$ and $\zeta'(0)$ for a $d$-dimensional ball as the base manifold of $\Omega$ in a few lower-dimensional cases.
Let us start with $d=2$, namely a $2$-dimensional base, meaning that $\Omega$ is of dimension $D=3$. In this case we have
\bea
\zeta_{D=3}(0)=-\frac{1}{48}-\frac{1}{16} \left(-1+4 \sigma ^2\right) \sin^{2}\theta_0\;,
\eea
and also
\bea
\zeta'_{D=3}(0)&=&-\frac{1}{24}\ln(\sin\theta_0)-\frac{1}{12}\ln\,\left(1+\cos\theta_0\right)+\frac{1}{2}\zeta'_{R}(-1)-\frac{3}{4}\zeta'_{R}(-2)-\frac{1}{2}\varphi(0)-\frac{11}{36}\frac{d^{3}\phi(x)}{dx^{3}}\Bigg|_{x=0}\nonumber\\
&-&\int_{0}^{\infty}du^{2}\ln u^{2}\frac{d}{du^{2}}a_{2}(\nu(u^{2}))-i\int_{0}^{\infty}\frac{\varphi(ix)+\varphi(-ix)}{e^{2\pi x}-1}dx+\frac{1}{6}\int_{0}^{\infty}dx\ln x\frac{d^{4}\phi(x)}{dx^{4}}\nonumber\\
&+&\int_{2}^{\infty}dx\,x^{-4}\phi(x)\;.
\eea
For $d=3$, i.e. $D=4$, we obtain
\bea
\zeta_{D=4}(0)&=&-\frac{1}{180}+\frac{1}{8} \left(-1+4 \sigma ^2\right) \sin^{2}\left(\frac{\theta_0}{2}\right)+\frac{1}{64} \left(-1+4 \sigma ^2\right) \left(-25+4 \sigma ^2\right) \sin^{4}\left(\frac{\theta_0}{2}\right)\nonumber\\
&-&\frac{1}{96} \left(-1+4 \sigma ^2\right) \left(-25+4 \sigma ^2\right)\sin^{6}\left(\frac{\theta_0}{2}\right)\;,
\eea
and
\bea
\zeta'_{D=4}(0)&=&\frac{1}{240}-\frac{1}{90}\ln(\sin\theta_0)+\frac{1}{120}\ln(1+\cos\theta_0)+\frac{1}{6}\zeta'_{R}(-1)-\frac{1}{2}\zeta'_{R}(-2)+\frac{1}{3}\zeta'_{R}(-3)-\frac{1}{2}\varphi(0)\nonumber\\
&-&\frac{25}{288}\frac{d^{4}\phi(x)}{dx^{4}}\Bigg|_{x=0}-\frac{1}{2}\int_{0}^{\infty}du^{2}\ln u^{2}\frac{d}{du^{2}}a_{3}(\nu(u^{2}))-i\int_{0}^{\infty}\frac{\varphi(ix)+\varphi(-ix)}{e^{2\pi x}-1}dx\nonumber\\
&+&\frac{1}{24}\int_{0}^{\infty}dx\ln x\frac{d^{5}\phi(x)}{dx^{5}}+\int_{1}^{\infty}dx\,x^{-5}\phi(x)\;.
\eea

For $d=4$, or equivalently $D=5$, one gets
\bea
\zeta_{D=5}(0)&=&\frac{17}{11520}-\frac{7}{192} \left(-1+4 \sigma ^2\right) \sin^{2}\left(\frac{\theta_0}{2}\right)-\frac{1}{192} \left(-1+4 \sigma ^2\right) \left(-59+16 \sigma ^2\right) \sin^{4}\left(\frac{\theta_0}{2}\right)\nonumber\\
&+&\frac{1}{24} \left(-1+4 \sigma ^2\right)\left(-13+4 \sigma ^2\right) \sin^{6}\left(\frac{\theta_0}{2}\right)-\frac{1}{48} \left(-1+4 \sigma ^2\right)\left(-13+4 \sigma ^2\right)\sin^{8}\left(\frac{\theta_0}{2}\right)\;,
\eea
and, finally,
\bea
\zeta'_{D=5}(0)&=&\frac{17}{5760}\ln(\sin\theta_0)+\frac{17}{2880}\ln(1+\cos\theta_0)-\frac{5}{64}\zeta'_{R}(-4)+\frac{7}{48}\zeta'_{R}(-3)-\frac{1}{32}\zeta'_{R}(-2)-\frac{1}{48}\zeta'_{R}(-1)\nonumber\\
&-&\frac{1}{2}\varphi(0)-\frac{137}{7200}\frac{d^{5}\phi(x)}{dx^{5}}\Bigg|_{x=0}+\frac{1}{24}\int_{0}^{\infty}du^{2}\ln u^{2}\frac{d}{du^{2}}\left[a_{2}(\nu(u^{2}))-4a_{4}(\nu(u^{2}))\right]\nonumber\\
&-&i\int_{0}^{\infty}\frac{\varphi(ix)+\varphi(-ix)}{e^{2\pi x}-1}dx+\frac{1}{120}\int_{0}^{\infty}dx\ln x\frac{d^{6}\phi(x)}{dx^{6}}+\int_{\frac{2}{3}}^{\infty}dx\,x^{-6}\phi(x)\;.
\eea

It is quite straightforward to see that the above results reproduce the known results for the ball once the limit for small $\theta_0$ is taken.
Moreover, for conformally invariant scalar fields it is not difficult to realize that the parameter $\sigma$ takes the value $1/2$. As one can easily see
from the above formulas, when this case is considered our results for $\zeta(0)$ reduces to the ones obtained in flat space as it should be expected.
We would like to stress, here, that although the integrals appearing in the above formulas for $\zeta'(0)$ are not explicitly computable in the general case, one could
start immediately a numerical analysis obtaining, in this way, very specific results.

\section{Concluding Remarks}
In this paper we have presented a technique for the evaluation of the zeta determinant for the massive Laplacian on a Riemann cap (or spherical suspension). Riemann caps are generalization of deformed spheres in the sense that the base manifold is compact and boundaryless, but otherwise general, and present a singular Riemannian structure analogous to that of generalized cones. In the spirit of ref.~\cite{cheeger83}, we expressed the functional determinant in terms of the base zeta function and found some general formulas for the analytical continuation of the zeta function and its derivative at $s=0$, which may be used to obtain the determinant and, in a more physical context, the effective action for a quantum field propagating on the above manifold. Finally, we specialized our general formulas to the case of a ball as a base manifold and presented the, still general, results in such case. Explicit formulas are finally given in $D=3,~4,$ and $5$ dimensions recovering previously obtained results.

Aside of presenting results which are more general than those available in the literature, one of the goals of our work was to illustrate a method to compute the zeta determinant alternative to that of ref.~\cite{bordag}, which is the one now commonly used in various situations. The basic idea behind the method of ref.~\cite{bordag} is to express the zeta function as an integral in a way similar to what we did here, and then add and subtract the asymptotic expansion of the integrand. This method of asymptotic subtraction allows a direct way of isolating the divergences and thus regularize the functional determinant by direct computation of the integral of the asymptotic part. In general, this last step depends on the actual behavior of the egeinfunctions and it is non-trivial. In fact, in the case of Riemann caps analyzed here, it does not seem to be the most convenient way to proceed. Alternatively, the method we have used here is based on an adaptation of a Lemma of analytic continuation, expressed through formulas (\ref{phix})-(\ref{continue}), that relates the zeta function and its derivative at $s=0$ to the coefficients of the (uniform) asymptotic expansion of the eigenfunctions. It goes without saying that it can also be used in the well known cases of spheres and balls, reproducing previous results.

One situation of physical interest, where the present technique may be used, is in the context of higher codimension brane-world models. In these models cone-type singularities arise and one standard way of regularization is to cut the tip of the cone, thus removing the singular behavior \cite{derahm,burgess,kobayashi,minamitsuji}. In fact, the case of conic manifolds has received some recent attention, and, for instance, ref.~\cite{kirsten06} describes some unusual and pathological behavior of the zeta function that cannot be easily recovered by simply taking the limit of cone regularized in the above way. One possible alternative is to smoothen the cone by rounding it with a spherical cap. In this way, recuperating the conical structure is rather transparent and can easily done by taking the limit of small angles, where the metric of the Riemann cap reduces to that of the cone.

Another advantage of the method used here is that the functional determinant is expressed in terms of the coefficients of the asymptotic expansion (\ref{phix}) making the method suitable for numerical implementations. This may be particularly useful in computation of the Casimir energy. In fact, the method utilized here can be easily generalized to consider the analytic continuation of the zeta function to other points in the complex plane allowing to: obtain the analytic continuation of the zeta to other values of $s$ and thus constrain further heat-kernel coefficients; develop a different way to compute the Casimir energy, which requires the value of the zeta determinant at $s=-1/2$.

The above issues are currently being considered and we hope to report on our progress soon.

\acknowledgements
The authors are grateful to Alexander Yu. Kamenshchick, Klaus Kirsten, Misao Sasaki and Takahiro Tanaka for very useful discussions. AF acknowledges the support of the Japanese Society for Promotion of Science through contracts N. 19GS0219 and N. 20740133. AF would like to express his gratitude to the Department of Mathematics at Baylor University for the kind hospitality during which this work was initiated.

\begin{appendix}

\section{Polynomials $a_{n}(\cos\theta_0)$}
For the reader's convenience, we list, here, the polynomials $a_{n}(\cos\theta_0)$ up to the fourth order. By utilizing the recurrence relations (\ref{2}) and (\ref{3})
and by subsequently taking the limit as $u\to 0$ one obtains
\bea
a_1(\sigma,\theta_0)=-\frac{1}{12}+\frac{1}{4} \left(1-4 \sigma ^2\right) \sin^{2}\left(\frac{\theta _0}{2}\right)\;,
\eea
\bea
a_2(\sigma,\theta_0)&=&-\frac{1}{4} \left(1-4 \sigma ^2\right)\sin^{2}\left(\frac{\theta _0}{2}\right)+\frac{1}{4} \left(1-4 \sigma ^2\right)\sin^{4}\left(\frac{\theta _0}{2}\right)\;,
\eea
\bea
a_3(\sigma,\theta_0)&=& \frac{1}{360} + \frac{1}{4} \left(1 - 4 \sigma^2\right) \sin^2\left(\frac{\theta_0}{2}\right) -
 \frac{1}{32} \left(25 - 104 \sigma^2 + 16 \sigma^4\right)\sin^4\left(\frac{\theta_0}{2}\right) \nonumber\\
 &+&\frac{1}{48} \left(25 - 104 \sigma^2 + 16 \sigma^4\right) \sin^6\left(\frac{\theta_0}{2}\right)\;,
\eea
\bea
a_4(\sigma,\theta_0)&=&-\frac{1}{4} \left(1-4 \sigma ^2\right) \sin^2\left(\frac{\theta_0}{2}\right)+\frac{1}{8} \left(15-64 \sigma ^2+16 \sigma ^4\right) \sin^4\left(\frac{\theta_0}{2}\right)\nonumber\\
&-&\frac{1}{4} \left(13-56 \sigma ^2+16 \sigma ^4\right) \sin^6\left(\frac{\theta_0}{2}\right)+\frac{1}{8} \left(13-56 \sigma ^2+16 \sigma ^4\right) \sin^8\left(\frac{\theta_0}{2}\right)\;.
\eea

\end{appendix}

\end{document}